# Reply to "Comment on 'Compositional and Microchemical Evidence of Piezonuclear Fission Reactions in Rock Specimens Subjected to Compression Tests' [Strain 47 (Suppl. 2), 282 (2011)]" by G. Amato et al.

A. Carpinteri[1,2], A. Chiodoni[3], A. Manuello[1,2], R. Sandrone[4]

[1]Department of Structural, Building and Geotechnical Engineering, Politecnico di Torino, Corso Duca degli Abruzzi 24, 10129 Torino, Italy.
[2]National Research Institute of Metrology, Strada delle Cacce 91, 10135 Torino, Italy.
[3]Center for Space Human Robotics@PoliTo, Istituto Italiano di Tecnologia, Corso Trento 21, 10129 Torino, Italy.
[4]Department of Environment, Land and Infrastructure Engineering, Politecnico di Torino, Corso Duca degli Abruzzi 24, 10129 Torino, Italy.

Amato et al.[1] maintain that certain data published in Carpinteri et al.[2] are not the result of independent measurements, and are in any case affected by uncertainties such that our conclusions are not valid. We will provide clarification regarding the validity of the disputed analyses, showing that they, like the others reported in the original article, are independent and that the uncertainties affecting them in no way invalidate our conclusions.

## I. INDEPENDENCE AND RELIABILITY

For the sake of clarity, the disputed analyses are referred in this paper as grouped in rows A through M, by Amato et al.[1] Some of these analyses were performed at different points of the same phengite lepidoblast (mica crystal in metamorphic rock). For the external surface, analyses 1-A, 11-A, 21-A, 28-A, 12-B, 22-B, 13-C, 4-D, 24-E, 27-H, 20-I were carried out on the same lepidoblast, whereas analyses 2-B, 3-C, 14-D, 9-D were carried out at different points of a second lepidoblast. For the fracture surface, 21-D, 24-D, 25-E, 26-F, 27-G, 20-H were also carried out on the same lepidoblast, as were analyses 23-J, 30-J, 19-J as well as analyses 15-K, 16-K. It should be emphasized that all of these analyses were performed on phengite lepidoblasts taken from a single specimen of less than one $dm^3$ in volume, rather than from specimens from the entire Luserna Stone outcrop area. Phengite composition can thus be regarded as particularly homogeneous from the chemical standpoint.

As indicated in the original paper, the investigations were carried out using EDS microprobe analysis[2,3]. The precision of these analyses depends on a number of factors, including the counting time, the quality of the standards, the condition of the sample surface, and the concentration of the measured elements. In any case, the good quality of all analyses published in Carpinteri et al.[2] was examined after measurement by calculating the atomic proportions using Minsort software[6]. This software calculates the elements' atomic proportions on the basis of the number of atoms of oxygen in the mineral formula (22 atoms for phengite and for biotite). The sum of the atoms of each element, multiplied by the element's valence, must saturate the valence of the oxygen atoms in all reliable analyses. The good quality of all the analyses is indicated by the atomic proportions shown in Tables 1 and 2 published in Carpinteri et al.[2]

The EDS analysis software used in this case provides percentage weight measurements of the element quantities with two decimals and a value corresponding to 1 sigma error[3]. By assuming that all oxygen present in the sample is in the form of oxides having known formulas, from the element percentages is possible to calculate the concentration of the different oxides, as is standard practice in mineralogy and petrography.

EDS measurements precision is in the order of magnitude of a few percent[4,5]. The minimum detection limits (MDLs) are typically less than 1%; and the dynamic range runs from the MDL to

100%, with a relative precision of 1% to 5% throughout the range[4].

When the measured percentage of the element is small (<1%), the error is restricted to the third decimal place, corresponding to 1% of the minimum detection limit. In this case, the third decimal place (as in most publications in the field) is not shown, since it is undoubtedly affected by instrumental error. This is the case of Mg and Ti, for which the conclusions made by Amato et al.[1] are incorrect. When the measured percentages are in the order of 1%, the second decimal place is a mere guard digit[7], since the precision is in the order of 0.01%. Similarly, for a measured percentage of 10%, uncertainty is in the order of 0.1%. In these cases, also the hundreds complement calculation of elements' concentrations can generate repeated data. For initial measured values that differ in the second decimal place, multiplying by factors, within a certain range, in order to calculate the hundreds complement, may lead to the same final value.

From the above considerations appears evident that the inference: "the chemical composition data published by Carpinteri et al.[2] cannot be the result of independent measurements" reported by Amato et al.[1] is incorrect, extremely forced and not sufficiently supported.

Finally, the use of mean values, as reported in Carpinteri et al.[2] does not detract from the published results' validity, as instrumental uncertainty is sufficiently below the deviation between compared mean values, as in the case of the Fe and Al concentrations shown in Figure 4 of Carpinteri et al.[2] The compositional variations in Fe and Al in phengite, in fact, involve the integer values of the percentage measurements, whereas instrumental error is in the order of some percents of the measured percentage[4,5].

## II. CONCLUSIONS

We believe that the additional information given above regarding the procedures used to measure and process phengite data provide further evidence of the compositional variations between external and fracture surfaces.

In addition, considering that all of the authors of Amato et al.[1] are engaged in research at the same institution as two of the authors of this reply, it is believed that the interests of scientific progress would be best served by collaborating in determining the repeatability of the experiments described in recent papers by Carpinteri et al. on the subject[2,8-12]. Further investigations, even if performed independently by other research groups, could lead to significant advances in understanding the physical phenomenon.

[1] G. Amato, G. Bertotti, O. Bottauscio, G. Crotti, F. Fiorillo, G. Mana, M. L. Rastello, P. Tavella, and F. Vinai, arXiv:1205.6418v1, (2012).
[2] A. Carpinteri, A. Chiodoni, A. Manuello, and R. Sandrone, Strain 47 (2), 282 (2011).
[3] Spectrum Synthesis – Accurate estimation of the Limit of Detection for any sample. INCA Energy, Oxford Instruments (2002) (http://www.oxford-instruments.com/incatips/tip7/index.html).
[4] Energy Dispersive X-Ray Microanalysis, Noran Instruments, Inc, (1999).
[5] ASTM E 1508-98, Standard Guide for Quantitative Analysis by EDS.
[6] K. Petrakakis and H. Dietrich– MINSORT: N. Jb. Miner. Mh., 8, 379 (1985).
[7] F. Pavese, arXiv:1206.2797v1, (2012).
[8] A. Carpinteri, F. Cardone, and G. Lacidogna, Strain 45, 332 (2009).
[9] F. Cardone, A. Carpinteri, and G. Lacidogna, Phys. Lett. A, 373, 4158 (2009).
[10] A. Carpinteri, F. Cardone, G. Lacidogna Exp. Mech., 50, 1235 (2010).
[11] A. Carpinteri, A. Manuello, A. Strain, 47(2), 267 (2011).
[12] A. Carpinteri, G. Lacidogna, A. Manuello, O. Borla, Rock Mech. and Rock Eng., 45(4), 445 (2012).